%% file: novel_approach_draft.tex
\documentclass[sigconf]{acmart}
\acmConference[EASE 2026]{The 30th International Conference on Evaluation and Assessment in Software Engineering}{9–12 June, 2026}{Glasgow, Scotland, United Kingdom}%%

\AtBeginDocument{%
  }

\usepackage{subcaption}

\usepackage{graphicx}
\usepackage[table]{xcolor}
\usepackage{colortbl}
\usepackage{float}
\usepackage{stfloats}
\usepackage{hyperref}
\usepackage[table]{xcolor}
\usepackage{adjustbox}
\usepackage{rotating}
\usepackage{amsmath}
\usepackage{url}
\usepackage{svg}
\usepackage{amsfonts}
\usepackage{listings}
\usepackage{float}
\newfloat{listing}{htbp}{lop}
\floatname{listing}{Listing}
\usepackage[most]{tcolorbox}
\usepackage{xcolor}
\usepackage{caption}

\newtcolorbox{rqbox}[1]{
  colback=blue!3,
  colframe=blue!50!black,
  title=\textbf{#1},
fonttitle=\small,
  boxrule=0.5pt,
  arc=1pt,
  left=3pt,
  right=3pt,
  top=2pt,
  bottom=2pt
}

\DeclareCaptionType{example}[Example][List of Examples]
\definecolor{codebg}{HTML}{F5F5F5}
\definecolor{framecol}{HTML}{4A90E2}

\begin{document}

\title{EnCoDe: Energy Estimation of Source Code At Design-Time}
% ==== AUTHORS ====
\author{Shailender Goyal}
\affiliation{
  \institution{Software Engineering Research Center, IIIT Hyderabad}
  \city{Hyderabad}
  \country{India}
}
\email{shailender.goyal@research.iiit.ac.in}

\author{Akhila Matathammal}
\affiliation{
  \institution{Software Engineering Research Center, IIIT Hyderabad}
  \city{Hyderabad}
  \country{India}
}
\email{akhila.matathammal@research.iiit.ac.in}

\author{Karthik Vaidhyanathan}
\affiliation{
  \institution{Software Engineering Research Center, IIIT Hyderabad}
  \city{Hyderabad}
  \country{India}
}
\email{karthik.vaidhyanathan@iiit.ac.in}





\begin{abstract}
\input{sections/1.Abstract}

\end{abstract}
\maketitle

\newenvironment{bluepar}{\color{blue}}{}

\input{sections/2.Introduction}

\input{sections/3.RelatedWorks}

\input{sections/Motivation}
\input{sections/5.Methodology}

\input{sections/6.ExperimentalSetup}
\input{sections/research_question}

\input{sections/7.Results}

\input{sections/8.Discussion}
\input{sections/9.Conclusion}

\section{Reproducibility Statement}
\label{sec:reproducibility}
All code and data used to produce the results in this paper is available in the following repository \footnote{\url{https://doi.org/10.5281/zenodo.18366913}}.
 % \begin{bluepar}
\section {Acknowledgements}
The authors would like to acknowledge the support of the ANRF Prime Minister Early Career Research Grant under the project SustAInd (ANRF/ECRG/2024/003379/ENS).
% \end{bluepar}
%% Bibliography
\bibliographystyle{ACM-Reference-Format}
\bibliography{acmart}





\end{document}

%% file: sections/1.Abstract.tex
Energy efficiency has emerged as a vital attribute of software quality, with significant implications for both environmental sustainability and operational costs. However, existing profiling tools operate only at runtime and coarse granularity, typically capturing energy at the process or method level. Such tools fail to expose how small code blocks, such as functions, loops, and conditionals, contribute to energy consumption, preventing developers from reasoning about and comparing the energy efficiency of programming constructs during design-time. 

To address this gap, we propose EnCoDe, a methodology for fine-grained, design-time energy estimation, with the following key contributions: (1) PowerLens, a novel measurement methodology that achieves reliable sub-millisecond energy readings for small code blocks; (2) Extensive empirical study on code blocks extracted from over 18,000 Python programs, uncovering linear and non-linear relationships between energy consumption and static code features such as structural, complexity, density, and contextual characteristics, resulting in a first-of-its-kind fine-grained dataset; and (3) Predictive modeling, in which machine learning models are trained on these features to accurately estimate and classify block-level energy consumption at design-time. Our results demonstrate stable, reproducible block-level estimations, with regressors achieving $R^2 = 0.75$ and classifiers achieving $80.6\%$ accuracy in identifying energy hotspots, enabling developers to localize and address inefficient code regions early in the development process without execution.

%% file: sections/2.Introduction.tex
\section{Introduction}

% 1. Context and the Problem
% 2. State of the Art and the Gap
% 3. Proposed approach and contributions
% 4. Paper Structure

%Industry analysts also increasingly recognize sustainability as a core software quality attribute: Gartner predicts that by 2027, around 30\% of large global enterprises will include software sustainability as a formal non-functional requirement in their development processes, up from less than 10 % in 2024

Source code is the fundamental component of software systems, defining the behavior of systems ranging from embedded IoT devices to global communication networks \cite{DiCosmo2017}. While hardware efficiency has improved significantly, the software running on it remains a massive consumer of energy, often written without regard for its environmental footprint due to software bloat and abstraction layers \cite{Wirth1995}. The aggregate energy consumption of inefficient code across billions of devices contributes heavily to global carbon emissions \cite{Belkhir2018, Freitag2021}. To mitigate this, we must move beyond traditional post-execution profiling and develop predictive models capable of estimating the energy consumption of code blocks \textbf{during the development phase}. By quantifying the energy cost of specific constructs before compilation, developers can treat energy as a tangible resource and make informed decisions in real-time, rather than relying solely on retrospective measurements \cite{Pang2016, Pinto2017}.  

\begin{comment}

Software systems are a fundamental component of modern infrastructure, powering everything from global communication networks to the rapidly expanding Internet of Things (IoT). However, this scale comes with a substantial environmental cost. While individual applications may seem negligible, their aggregate energy consumption is massive. Data centers alone account for a significant fraction of global electrical usage \cite{Marantos2023TSUSC, doi:10.1126/science.aba3758}, and the cumulative energy consumption of software on billions of consumer devices contributes heavily to carbon emissions. Consequently, energy efficiency is no longer just a requirement for high-performance computing; it is a necessity for all general-purpose software. Reducing energy consumption helps lower the environmental carbon footprint and significantly reduces operational costs for companies running software at scale\cite{10.1145/2597073.2597097}. 

\end{comment}

Despite this need, developing energy-efficient software remains structurally difficult. Current assessment techniques rely predominantly on \textit{runtime measurements} using hardware power meters or processor-level counters like Intel RAPL \cite{10.1145/2425248.2425252}. This creates a reactive \textit{"measure-after-build"} workflow, where a program must be fully implemented and executed before its energy behavior can be observed. This late-stage feedback is costly; discovering an energy defect after deployment often requires expensive refactoring cycles. Furthermore, existing profiling tools often provide insufficient granularity for effective optimization \cite{7429297}. They typically report energy usage at the process or application level, failing to attribute computation to specific code constructs such as loops or conditional blocks that developers actively manage \cite{Khan2018RAPLinAction, Shivadharshan2024CPPJoules}. This disconnect leaves developers unable to pinpoint the exact source of energy inefficiencies, reducing optimization efforts to trial and error \cite{Pinto2017}. 

To address these limitations, proactive approaches are needed that shift energy assessment from a runtime concern to a design-time quality attribute. Developers currently rely on static analysis tools and linters such as \textit{SonarQube} \cite{Lenarduzzi2020JSS}, \textit{PMD} \cite{AlOmar2023SEET_PMD},\textit{CheckStyle}\footnote{\texttt{Checkstyle}: \url{https://checkstyle.sourceforge.io/}}
, and \textit{SpotBugs}\footnote{\texttt{SpotBugs}: \url{https://spotbugs.github.io/}} to identify bugs and enforce coding standards before code is even compiled. A similar proactive approach is required for energy efficiency. However, there are insufficient tools that can estimate the energy implications of source code structure during the development phase. To the best of our knowledge, no existing approach provides fine-grained, block-level energy estimation purely from source code at design time without requiring code execution or specialized hardware profiling.  This gap forces energy efficiency to remain a retroactive afterthought rather than a proactive design parameter. 

In this paper, we propose \textbf{\textit{EnCoDe}}, a novel methodology for \textbf{\textit{"design-time, fine-grained estimation of energy consumption"}}. Our methodology bridges the gap between static code structure and dynamic energy behavior. To achieve this, we first address the measurement gap by developing \textbf{\textit{PowerLens}}, a custom measurement methodology capable of reliably profiling code blocks at sub-millisecond granularity. This infrastructure allows us to construct a ground-truth dataset of code blocks annotated with precise energy consumption values. Leveraging this dataset, we extract rich structural features from \textbf{\textit{Abstract Syntax Trees (ASTs)}} and train machine learning models to predict block-level energy consumption statically. We explicitly employ classical, low-overhead machine learning approaches (e.g., Random Forests, Gradient Boosting) rather than energy-intensive deep learning models. This alignment with the principles of Data-Centric Green AI~\cite{Verdecchia2022DataCentricGreenAI} ensures that our solution does not ironically consume excessive energy in the pursuit of energy efficiency, maintaining a low carbon footprint for the tool itself. 

The specific contributions of this paper are as follows:
\begin{enumerate}
    \item \textbf{\textit{PowerLens Measurement:}} We present a novel measurement methodology that utilizes execution amplification and temporal synchronization to \textbf{achieve reliable sub-millisecond energy readings}. This allows us to accurately profile microsecond-scale code blocks that are otherwise invisible to standard hardware counters.
    \item \textbf{\textit{Fine-Grained Energy Dataset:}} We conducted an extensive empirical study on executable code blocks extracted from over \textbf{18,000 Python programs}. This study uncovers the linear and non-linear relationships between static code features such as structural complexity, operator density, and energy consumption.
    \item \textbf{\textit{Predictive Modeling and Validation:}} We demonstrate that classical machine learning models trained on these static features can accurately estimate energy consumption at development time. Our results show that EnCoDe achieves a coefficient of determination \textbf{\(R^2\) of 0.75} for regression and an \textbf{accuracy of 80.6\%} for classifying energy hotspots, enabling developers to identify and address inefficiencies early in the software development lifecycle. 
\end{enumerate}

The remainder of this paper is structured as follows. \textbf{Section ~\ref{sec:related}} reviews the related work in software energy measurement and static analysis. \textbf{Section ~\ref{sec:methodology}} detailed the EnCoDe methodology, including the PowerLens methodology and the feature extraction process. \textbf{Section ~\ref{sec:exp-setup}} describes the experimental setup and dataset construction. \textbf{Section ~\ref{sec:research-questions}} formulates the research questions guiding this study. \textbf{Section ~\ref{sec:results}} presents our experimental results and analysis. \textbf{Section ~\ref{sec:Discussion}} discusses the implications of our finding.\textbf{section ~\ref{sec:threats}} addresses the threats to validity. Finally, \textbf{Section ~\ref{sec:conclusion}} documents our conclusions and outlines future work.

%% file: sections/3.RelatedWorks.tex
\section{Related Work}
\label{sec:related}
Research into software energy efficiency has primarily focused on two distinct areas: runtime measurement infrastructures and predictive modeling for specific domains. 

\subsection{Software Energy Measurement}

Accurate energy measurement is the prerequisite for optimization. The most reliable method involves external hardware power meters, but their high cost and non-scalability limit their use to lab environments \cite{Hindle2014GreenMining}. Consequently, the research community has shifted toward software-base power models. Intel's RAPL (Running Average Power Limit) has become the de facto standard for accessing on-chip energy counters \cite{Khan2018RAPLinAction}. However, as noted in recent studies, RAPL's update frequency (approx. 1ms) is often too coarse to capture the energy consumption of short-lived code constructs \cite{10.1145/2425248.2425252}. 
To address this granularity issue, tools like ALEA \cite{7429297} and finer-grained profilers for embedded systems has been proposed. These approaches often use probabilistic sampling or instruction-level characterization to attribute energy to specific code regions. 
% {\color{red}While effective for post-hoc analysis, they remain runtime techniques: they require the code to be fully executable and the hardware to be available, preventing their use during the early design or coding phases.}

\subsection{Static Analysis and Energy Prediction}

To move energy assessment earlier in the lifecycle, researchers have explored static analysis and machine learning. 
In the mobile domain, approaches like \textit{MLEE}\cite{ALVI2021100594} and \textit{EcoAndroid} \cite{Ribeiro2021EcoAndroid}, have successfully demonstrated that structural metrics can predict energy consumption for Android applications. Similarly, \textit{FlipFlop} \cite{Sharma2024FlipFlop} utilizes static analysis to optimize GPU kernel configurations for AI workloads. 

For general-purpose languages like Python, tools such as \textit{GreenPy} \cite{Reya2023GreenPyEA} provide application-level energy insights. However, these existing approaches typically operate at the level of entire applications, methods, or specific API calls. They lack the resolution to estimate the energy cost of fundamental control flow blocks (e.g., individual \texttt{for} loops or \texttt{if} conditions) in general-purpose context.  Furthermore, many existing models can rely on deep learning, which can itself be energy-intensive. By contrast, EnCoDe targets this specific granularity gap using lightweight, interpretable models compliant with Green AI principles.

%% file: sections/Motivation.tex
\begin{figure}[H]
  \centering
  \includegraphics[width=\linewidth]{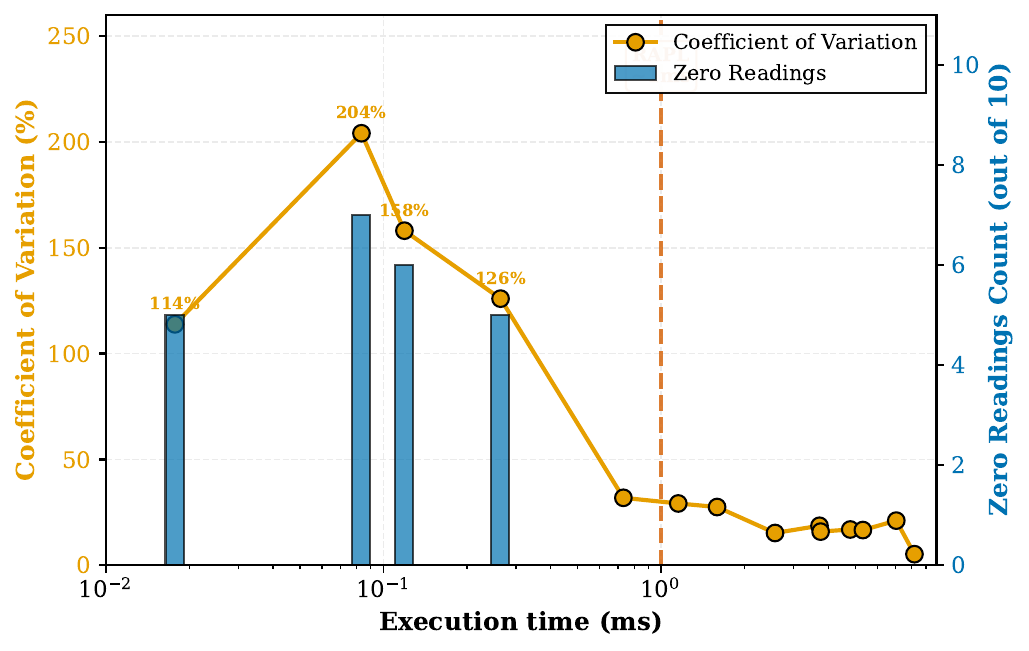} 
  \caption{Quality of RAPL's Measurements over Execution Time of the Workload (ms in Log scale)}
  \label{fig:rapl-quality}
\end{figure}
\section{Motivation}
\label{sec:motivation}
% 1. The Core Challenge in detail 
% A Motivation Example

Developing energy efficient software would require reliable insight into how individual fragments of source code contribute to energy consumption. Existing measurement techniques use hardware based counters such as Intel's Running Average Power Limit (RAPL). While effective for coarse grained profiling, but RAPL's millisecond resolution is too large for such code fragments. As Figure~\ref{fig:rapl-quality} demonstrates that workloads under 1ms register more than 110\% variation and 4-5 runs get zero readings (out of 10 runs). Even above 1ms, the variation is significant. Although hardware power meters provide higher resolution, they are severely limited in terms of accessibility, reproducibility, and usability in distributed systems. To address this fundamental limitation, we developed \textbf{PowerLens}, a methodology for obtaining precise and reproducible energy readings at sub-millisecond granularity. However, accurate measurements alone are insufficient; developers need energy information before code execution. To enable this, we propose \textbf{EnCoDe}, which trains machine learning models on PowerLens measurements to predict block-level energy consumption directly from source code features, shifting energy assessment from a runtime profiling concern to a proactive design-time estimation.

 % \cite{ALVI2021100594} \cite{inproceedings} \cite{10.1145/2425248.2425252}

\begin{comment}

\begin{figure*}[htbp]
  \centering
  \includegraphics[width=0.95\textwidth]{figures/EnCoDe.png}
  \caption{Design Time Energy Estimation Framework}
  \label{fig:framework}
\end{figure*}

\end{comment}

%% file: sections/5.Methodology.tex
\section{Methodology}
\label{sec:methodology}

% Architectural Overview. Begin with High-Level Diagram and then briefly walk into the main stages. 
% Implementation and Experimental Setup is seperate section. Introduce the Research Questions here. 

This section details the methodology leveraged by \textit{EnCoDe}. It is organized into a sequence of phases, beginning with the creation of a ground-truth dataset via our novel \textit{PowerLens} measurement methodology, followed by feature extraction, and culminating in the training of predictive models. 

To further explain the methodology followed in detail, we will use the following simple running example of a  Python function (\textit{Listing \ref{lst:calculate_score}}), demonstrating how it is transformed from raw code into a final energy estimate.

\begin{listing}[ht]
\centering
\begin{tcolorbox}[
    colback=codebg,
    colframe=framecol,
    arc=5mm,
    boxrule=1pt,
    title=\textbf{Example: \texttt{score\_function.py}},
    coltitle=black,
    fonttitle=\bfseries,
    top=1mm, bottom=1mm, left=2mm, right=2mm
]
\begin{lstlisting}[language=Python, basicstyle=\small\ttfamily, breaklines=true]
def calculate_score(items, threshold):
    score = 0
    for item in items:
        if item > threshold:
            score += (item * 2)
    return score
\end{lstlisting}
\end{tcolorbox}
\caption{Example: score computation function in Python}
\label{lst:calculate_score}
\end{listing}

\begin{figure*}[htbp]
  \centering
  \includegraphics[width=0.95\textwidth]{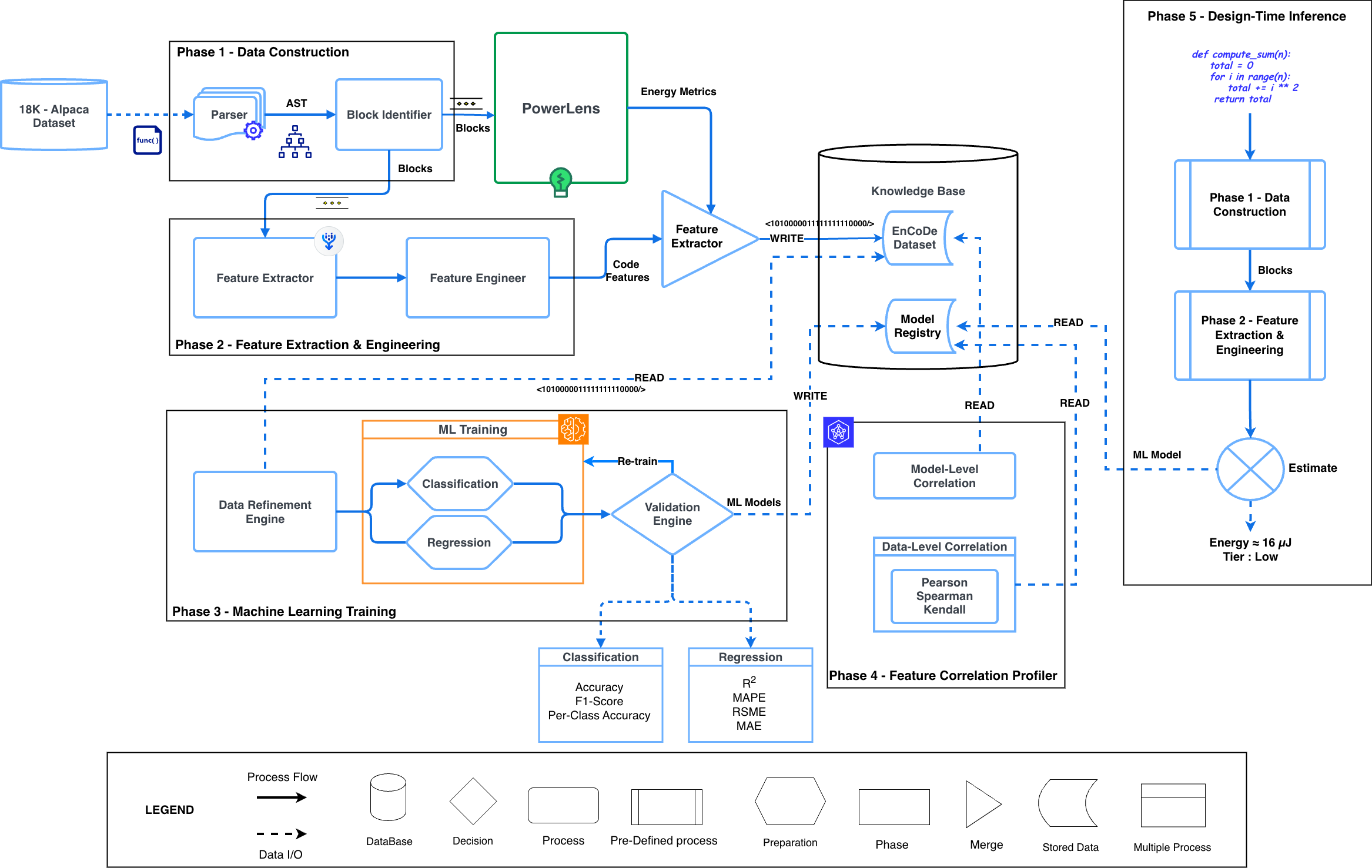}
  \caption{Design Time Energy Estimation Methodology}
  \label{fig:framework}
\end{figure*}

% This function contains several distinct computational units (blocks) such as a function definition (\textit{def}), a loop (\textit{for}), and a conditional (\textit{if}), whose individual energy contributions are precisely what \textit{EnCoDe} is designed to analyze. 

\subsection{Phase 1 - Data Construction}
\begin{figure*}[htbp]
  \centering
  \includegraphics[width=0.95\textwidth]{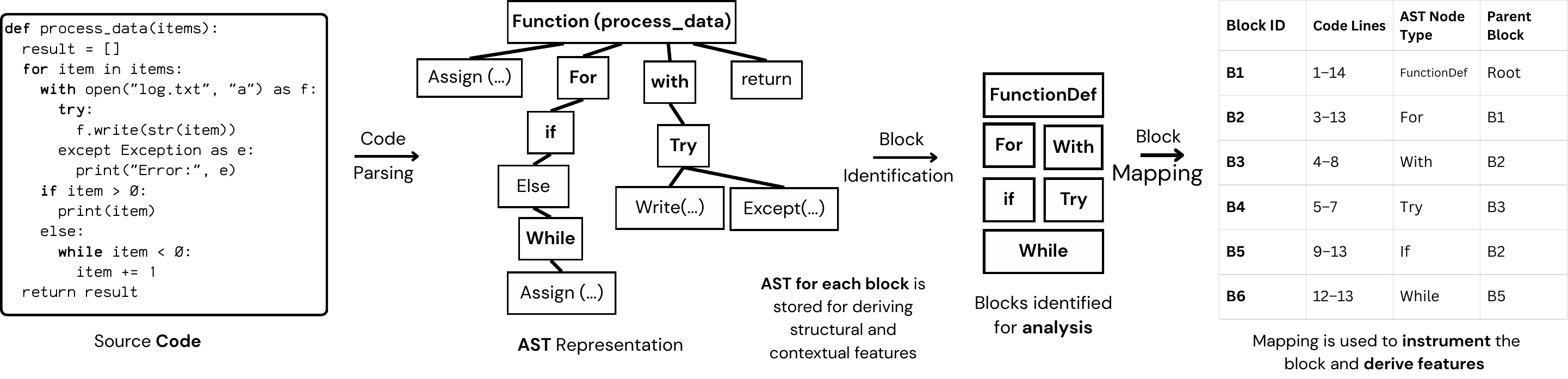}
  \caption{Code Parsing to identify blocks from AST}
  \label{fig:block_identification}
\end{figure*}

\noindent
The first phase of our methodology, as shown in \textit{Figure \ref{fig:framework}}, transforms a raw program text into a structured, analyzable computational block. This is achieved by first parsing the source code \(\mathcal{P}\) to its \textbf{Abstract Syntax Tree (AST)}, a hierarchical representation of the code's structure\cite{Aho1986CompilersPT, sun2023abstractsyntaxtreeprogramming}, as shown in \textit{Figure \ref{fig:block_identification}}. We then traverse the AST to identify node types that represent the root of distinct blocks. We define these block-rooting node types as:

\[ \mathcal{V}_{block} = \{FunctionDef, For, While, If, Try, With\} \]

Each node of these types, along with the entire subtree rooted at that node, is designated as a distinct block, as shown in \textit{Figure \ref{fig:block_identification}}. 

% For example, after parsing our \textit{Listing \ref{lst:calculate_score}} into its AST, this phase identifies all nodes whose types match our defined set of block-rooting nodes (\( \mathcal{V}_{block} \)). This process yields three distinct, nested blocks, each corresponding to a specific AST subtree:

% \begin{itemize}
%     \item \textbf{B0}: Root node of the AST (module level)
%       \begin{itemize}
%         \item \textbf{B1}: \textit{FunctionDef}(\texttt{calculate\_score})
%           \begin{itemize}
%             \item \textbf{B2}: \textit{For} loop inside the function
%               \begin{itemize}
%                 \item \textbf{B3}: \textit{If} statement inside the loop
%               \end{itemize}
%           \end{itemize}
%       \end{itemize}
% \end{itemize}

For \textit{Listing~\ref{lst:calculate_score}}, this yields three nested blocks: B1 (\textit{FunctionDef} \texttt{calculate\_score}), B2 (\textit{For} loop), and B3 (\textit{If} statement inside B2), preserving their containment hierarchy.

The output of this phase is a collection of AST subtrees, each treated as an atomic block. Crucially, their hierarchical context is preserved (e.g., we know B3 is nested within B2), which is essential for all subsequent measurement and feature extraction.

\begin{figure*}[htbp]
  \centering
  \includegraphics[width=0.95\textwidth]{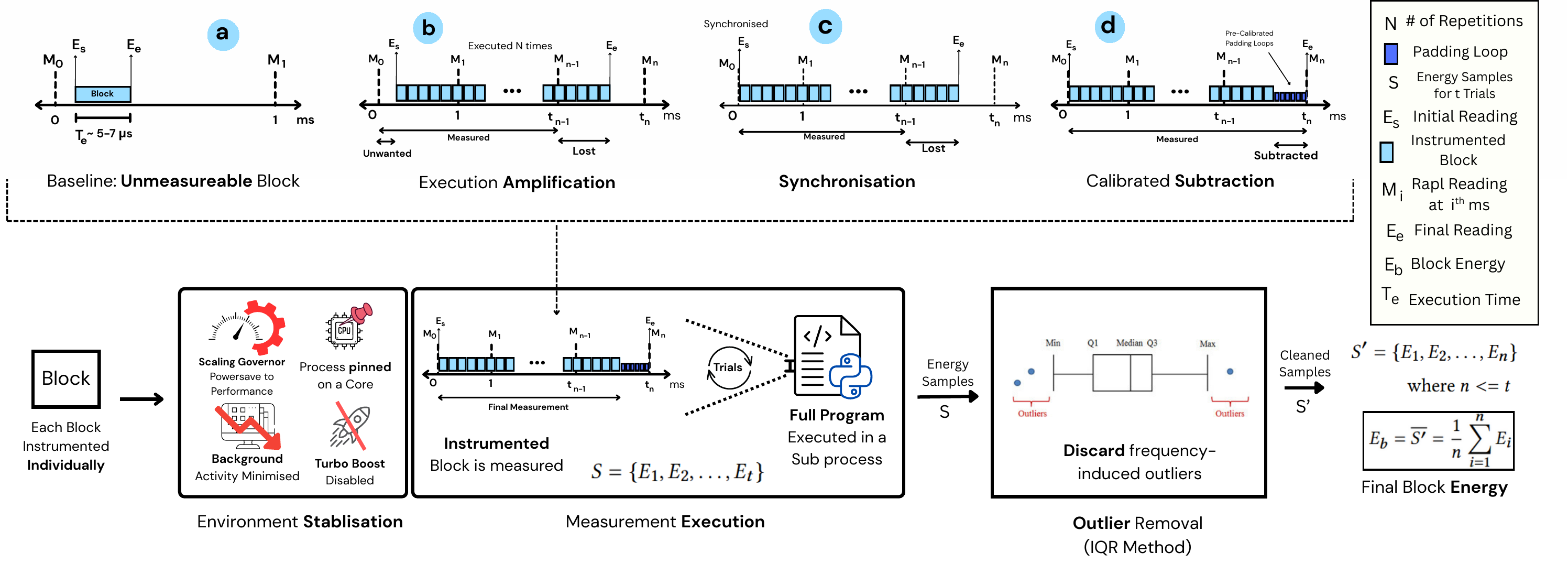}
  \caption{PowerLens : Sub-Millisecond Energy Measurement Methodology}
  \label{fig:powerlens}
\end{figure*}
% \begin{figure}[t]
%   \centering
%   \includegraphics[width=\linewidth]{figures/block_identification.pdf}
%   \caption{Code Parsing to identify blocks from AST}
%   \label{fig:block_identification}
% \end{figure}
% \begin{figure}[t]
%   \centering
%   \includegraphics[width=\linewidth]{figures/powerlens.pdf}
%   \caption{PowerLens: Sub-Millisecond Energy Measurement Methodology}
%   \label{fig:powerlens}
% \end{figure}

\subsection{PowerLens: Fine-Grained Energy Measurement}

With the code blocks identified, \textit{PowerLens} measures their energy consumption to create our ground-truth dataset. This is the core technical challenge of our work: a block like the \textit{if} statement (B3) as mentioned in our phase 1, executes in microseconds, far too fast for standard hardware counters like Intel's RAPL to measure directly \cite{ 10.1145/2425248.2425252}. Existing tools, therefore, cannot provide this data, as discussed in related work. \textit{PowerLens} addresses this fundamental measurement 
problem by extending prior work~\cite{10.1145/2425248.2425252}, operating 
through four coordinated mechanisms.

\subsubsection{Environmental Stabilization}
\label{sec:enviromental_Stab}
To ensure measurements are stable, we must isolate the execution from system “noise”. Fluctuations in CPU frequency or background processes can easily distort the tiny energy signature. \textit{PowerLens} enforces a stable environment by setting frequency scaling governor to performance, disabling Turbo Boost, pinning the measurement process to a single core, and suspending non-essential background processes. This ensures that the energy we measure is attributable to the code block itself.

\subsubsection{Execution Amplification.}

A single execution of a block like our \textit{if} statement (B3) is too fast to register on RAPL's millisecond-scale counters. To make this imperceptible signal measurable, \textit{PowerLens} wraps the target block in a tight loop and executes it \( N \) times. This amplifies the total energy consumed to a level that rises clearly above the millisecond-scale. The total energy \( E_{total} \), is calculated by taking two snapshots of the RAPL energy counter:\( (E_s) \) taken before the execution window, and \( (E_e) \) taken immediately after, as shown in \textit{Figure \ref{fig:powerlens}}. The measured difference is modeled as: 

\[ E_{total}(b, N) = E_e - E_s = N * E(b) + \epsilon_N \]

where \( E_b \) is the true energy of a single execution and \( \epsilon_N \) is the measurement noise which becomes negligible when averaged over N repetitions. By linear aggregation, the estimated mean energy per execution, \( \hat{E(b)} \), becomes:

\[ \hat{E(b)} = \frac{E_{total}(b,N)}{N} \]

For example, to measure B3 as discussed in Phase 1, we execute it N times. If the total measured energy \( E_{total} \) is \( 0.04 \) Joules (J), our initial measurement for a single execution is \( \hat{E}(B3) = 0.04 J / 1000 = 4.0 * 10^{-5} J \). 

\subsubsection{Synchronization - Aligning Execution and Sampling}

Amplification alone is insufficient. If the execution loop starts or end in the middle of a RAPL sampling window, the measurement will be corrupted by energy from unrelated computations, as illustrated by the “unwanted” energy in \textit{Figure \ref{fig:powerlens}}. To solve this, \textit{PowerLens} synchronizes the start of the execution loop with the RAPL counter's refresh cycle.  The start time \( t_0 \) is formally defined as the first moment a counter update is detected after the previous measurement start \( t_s \):

\[ t_0 = min\{ t > t_s | M_{i+1} > M_i \} \]

where \( M_i \) is the cumulative RAPL reading time at time tick \textit{i}. This precise alignment guarantees that the captured energy is attributable only to our target block.

\subsubsection{Calibrated Subtraction}

As amplified execution rarely finishes exactly on a RAPL boundary, a small “padding” interval is needed to complete the measurement window, and this padding consumes energy that must be removed. \textit{PowerLens} subtracts a pre-calibrated energy cost for this padding, \( E_{pad}(\delta t) \), which is modeled as a linear function of the padding duration \( \delta t \). The final, net energy for a single trial is then:

\[ E_{net}(b) = \frac{E_{total}(b,N) - E_{pad}(\delta t)}{N} \]

\subsubsection{Statistical Aggregation - Achieving Reproducibility}

To ensure the final value is robust against any residual system noise (eg., transient interrupts, SMM), the entire process is repeated \(m (=10)\) times. Outliers are filtered from these trials using the standard \textit{Inter-quartile Range (IQR)} rule \cite{alma9995263502466}. The final energy signature for a block, \( \hat{E(b)} \), is  the mean of the stable, non-outlier observations:

\[ \bar{E}(b) = \frac{1}{|\Omega(b)|} \sum_{i \in \Omega(b)} E_{net}^{(i)}(b) \]

where \(\Omega(b)\) denotes the non-outlier trial set. 

This is applied independently to each block, producing reliable, ground-truth energy values: \(\{B1: 7.7e-2 J, B2: 7.5e-2 J, B3: 1.4e-6 J\}\). This data, which could not be obtained with prior software methods forms our ground truth.

\subsection{Phase 2 - Feature Extraction and Engineering}

After energy measurements, the next step is to create a numerical "fingerprint" for each block based on its source code\cite{10.1145/3639475.3640110}. The goal is to compute a set of static features that capture the block's structural and syntactic characteristics, which we hypothesize are predictive of its energy consumption. This process creates the foundation for our predictive models, as illustrated in \textit{Figure \ref{fig:framework}}.

\subsubsection{Feature Extractor}

This component operates on the AST of each block, computing a feature vector \( x_b \) composed of 33 metrics grouped into seven categories. These categories are designed to capture different facets of code: 

% \smallskip
%\begin{itemize}
\noindent \textbf{Basic Metrics (5)}: Capture the block's overall scale (e.g., AST node count, depth). These serve as a baseline proxy for the total number of components.

% \smallskip
\noindent
\textbf{Complexity Metrics (4)}: Quantify logical intricacy using measures like cyclomatic complexity\cite{1702388}. Prior studies have established a direct correlation between high complexity and increased power consumption, as it often translates to more conditional branches in the instruction stream \cite{Keong2015MetricsPower}.

% \smallskip
\noindent
\textbf{Density Metrics (5)}: Measure the concentration of computational work, such as the ratio of operators to total AST nodes. Dense code regions often correspond to higher instruction throughput and sustained CPU activity.

% \smallskip
\noindent
\textbf{Diversity Metrics (6)}: Assess the heterogeneity of code elements (e.g., entropy of operator types). High diversity may engage a wider range of CPU functional units, influencing micro-architectural energy states.

% \smallskip
\noindent
\textbf{Structural Metrics (3)}: Capture the shape of the control-flow graph (e.g., branching factor), which impacts instruction fetch and branch prediction energy costs. \cite{10.1145/3212695}

% \smallskip
\noindent
\textbf{Code Pattern Metrics (5)}: Identify the presence of semantic constructs like loops and conditionals, which are primary determinants of a block's dynamic execution behavior.

% \smallskip
\noindent
\textbf{Halstead Metrics (5)}: Model the informational complexity of code (e.g., program volume, program effort)\cite{10.5555/540137}. While designed for cognitive effort, Halstead metrics are considered key indicators of software maintainability, a crucial aspect of overall software sustainability and energy efficiency. 

% For example, the \textit{if} statement (B3) is converted into a feature vector. Example
% \emph{ASTDepth} (3),
% \emph{Cyclomatic Complexity} (2),
% \emph{Operator Density} (0.4),
% and a binary flag \emph{HasConditional} (1).

\subsubsection{Feature Integrator}

\noindent
This final component unifies the outputs of the previous phases to produce the EnCoDe Dataset, a key research artifact of this work, as illustrated in \textit{Figure \ref{fig:framework}}. The integrator aligns the feature vector \(x_b\) with the corresponding energy label \(\hat{E}(b)\) for every block.

% \noindent
% For example, the feature vector for the if statement (B3) is paired with its measured energy value Ē(B3) to create a single data point in our dataset: \((x_B3, 1.4e-6 J)\). 

\subsection{Phase 3 - Machine Learning Training}

\noindent
With the finalized EnCoDe dataset, the objective of this phase is to train predictive models that can learn the relationships between a block's static features and its measured energy consumption. We frame this as a supervised learning problem, training two complementary models:
  
\noindent
\textbf{Regression Model \((M_r)\):} This model learns to predict the continuous energy value (in Joules). This allows to precisely compare the expected energy cost of different implementations.

\noindent
\textbf{Classification Model \((M_c)\):} This model provides more intuitive classification into energy tiers. The ground-truth energy values are discretized into three tiers ("Low," "Medium," "High") using equal frequency binning. This provides "lint-like" warning to quickly draw attention to potential energy hotspots. 
% \begin{bluepar}
The classifier's architecture is agnostic to threshold placement,
practitioners can re-bin the training data to match their energy budget
constraints without retraining the feature extraction pipeline.
% \end{bluepar}
\noindent
 To ensure robustness and prevent bias from the skewed energy distribution, models are trained using stratified k-fold cross-validation and checked for overfitting.

\noindent
\textit{For our example:} The data points for our three blocks: \((x_{B1}, 7.7e-2 J)\), \((x_{B2}, 7.5e-2 J)\), and \((x_{B3}, 1.4e-6 J)\), are fed into the training process. The models learn, for instance, that feature vectors with \(HasLoop=1\) and high \(OperatorDensity\) (like \(x_{B2}\)) tend to have higher energy values, while simple conditional blocks (like \(x_{B3}\)) have very low energy.

\subsection{Knowledge Base}
\noindent
The Knowledge Base is the conceptual component in our methodology that stores analytical artifacts, decoupling resource-intensive training from real-time inference.
\noindent
The trained models and the full dataset are stored in the \textbf{Knowledge Base} (Fig.~\ref{fig:framework}), which holds the EnCoDe Dataset \((D_{EnCoDe})\) and a Model Registry containing \(M_r\) and \(M_c\), decoupling resource-intensive training from real-time inference.

\noindent
\textit{For our example:} The data point \((x_B3, 1.4e-6 J)\) for our if statement, along with the data for B1 and B2, is stored in the dataset. The trained models, \(M_r\) and \(M_c\), which have learned from these and thousands of other examples, are stored in the registry.

\subsection{Phase 4 - Feature Correlation Profiler}
\noindent
% A predictive model, even an accurate one, is of limited use if it functions as an unexplainable "black box." To trust our models and to provide developers with actionable insights, we must understand which code characteristics are the most significant drivers of energy consumption. 
The purpose of this phase is to provide crucial interpretability of the prediction models.
\noindent
To achieve this, we perform a dual analysis that cross-validates our findings: one perspective is derived from the trained model's behavior, and the other from direct statistical evidence in the data.
\noindent
\textbf{Model-Based Feature Importance:} First, we analyze the trained models to rank features based on how much they contribute to prediction accuracy. For tree-based ensembles, a feature's influence \(I(f_j)\) can be quantified as its average marginal reduction in prediction error across all decision trees in the model :
\[ I(f_j) = \mathbb{E}_{f \in \mathcal{F}}[\Delta Err(f_j)] \]

\noindent
\textbf{Data-Level Correlation:} Second, we perform a model-agnostic statistical analysis where we compute the correlation coefficients between  features and the measured energy values to capture different types of relationships: Pearson \((\rho_p)\) for linear \cite{1896RSPTA.187..253P}, Spearman \((\rho_s)\) for monotonic \cite{spearman04}, and Kendall \((\rho_k)\) for rank-based associations\cite{665905b2-6123-3642-832e-05dbc1f48979}. This collection of correlations is represented as:
\[ \mathcal{C} = \{ (f_j, \rho_p(f_j, E)), (f_j, \rho_s(f_j, E)), (f_j, \rho_k(f_j, E)) \} \]

\noindent
By confirming that both analyses highlight the same key features, we increase our confidence that the models are capturing genuine, underlying relationships between code structure and energy, rather than learning from spurious artifacts.

\noindent
\textit{For our example:} This dual analysis provides deep insights.

\noindent
For the \textit{for} loop (B2), the feature importance might rank the binary feature \(HasLoop=1\) and the \(CyclomaticComplexity\) metric very highly.
\noindent
Simultaneously, the data-level analysis would likely show a strong positive Spearman correlation \((\rho_s)\) between these same features and energy across the entire dataset.
\noindent
This convergence gives us high confidence that the model has correctly learned a fundamental truth: loops and complex logic are significant drivers of energy consumption. The profiler's output is a unified ranking of these influential features, bridging the gap between prediction and explanation.

\subsection{Phase 5 - Design-Time Inference}
\noindent
This phase operationalizes EnCoDe's goal of design-time energy estimation by translating static code blocks into predictions using pre-trained models.
\noindent
The inference engine takes a newly written or modified code block as input and performs the following steps, as in \textit{Figure \ref{fig:framework}} :

\noindent
\textbf{Block Identification and Feature Extraction:} The unseen code is parsed into its AST, and its constituent blocks are identified using the same mechanism as in Phase 1. The Feature Extractor from Phase 2 is then reused to compute the corresponding feature vector, \(x_b^{new}\), for each new block.

\noindent
\textbf{Model-Based Prediction:} The pre-trained regression (\(M_r)\) and classification \((M_c)\) models are retrieved from the Model Registry in the Knowledge Base by the inference engine to generate the predictions.

\noindent
Formally, the inference process can be represented as a transformation function \(\Phi\) that maps the new block \(b_{new}\) and the stored models to an energy estimate and a tier : 
% \[Φ: (b_{new}, M_r, M_c) → (\hat{E}_b, T_b)\]
\begin{equation}
\Phi \colon (b_{\text{new}}, M_r, M_c) \to (\hat{E}_b, T_b)
\end{equation}

\noindent
where \(\hat{E}_b\) is the predicted energy in Joules and \(T_b\) is the predicted tier (e.g., "Low," "Medium," "High"). Crucially, this entire process is static and does not require any code execution. 

\noindent
\textit{For our example:} Imagine a developer is refactoring our Listing \ref{lst:calculate_score} function and replaces the for loop (B2) with a while loop.
 As they finish writing the new while loop block, energy is inferred again. The regression model \(M_r\) might predict an energy value of \(\hat{E}_b = 6.9e-2 J\). Simultaneously, the classification model \(M_c\) might predict the tier \(T_b\) = "Medium".
% \noindent
% This feedback, either the precise numerical estimate or the intuitive "Medium" tier warning, can then be displayed directly in the developer's editor, fulfilling the goal of design-time energy awareness.

%% file: sections/6.ExperimentalSetup.tex
\section{Experimental Setup}
\label{sec:exp-setup}
All experiments were executed on a dedicated, isolated machine under controlled conditions to maximize measurement stability and reproducibility.

\textbf{Hardware and system environment.}
Experiments were performed on an Intel CPU (model: \texttt{i7-6700K}; with 16\,GB DDR4 RAM, running Pop!\_OS 22.04 (64 bit). To reduce measurement noise, we stabilized the environment as in Section \ref{sec:enviromental_Stab}. We loaded the \texttt{msr} kernel module to access Model-Specific Registers for RAPL readings.

\textbf{Software environment.}
All experiments used Python~3.12. Block extraction relied on the standard \texttt{ast} module and custom instrumentation utilities. All code, experiment scripts, and seed values for pseudo-random operations are archived and available in the artifact repository. We set a fixed random seed for all learning experiments to ensure reproducible train/test splits and model initializations.

\textbf{Dataset.}
The dataset was constructed from a larger corpus of Python programs\footnote{\url{https://huggingface.co/datasets/iamtarun/python_code_instructions_18k_alpaca}}. After parsing with \texttt{ast}, we extracted executable code blocks defined as function bodies, loop bodies (\texttt{for}, \texttt{while}), conditional bodies (\texttt{if}), exception handlers (\texttt{try/except}), and context manager blocks (\texttt{with}). We filtered out non-executable blocks as we need ground truth measurements. From \textbf{18,612} code files, we extract \textbf{14,000+} executable blocks, retaining \textbf{8,000+} that exceed 1 µs (measurable with N=1000). Each block was annotated with a set of \textbf{33 static features} comprising of AST derived features and standard code properties.

\textbf{Measurement protocol.}
Energy readings were taken from processor energy counters exposed via Intel RAPL through MSR access. Specifically, we read the \texttt{PACKAGE} domains using the MSR interface. All MSR reads were executed with root privileges and synchronized with our measurement control loop to avoid partial-window reads. 
Block execution times are typically in the microsecond range, while RAPL updates at millisecond granularity; to bridge this mismatch we used deterministic amplification and synchronization as follows:

\textbf{Evaluation protocol and statistical analysis.}
We evaluated both regression and classification formulations. Regression performance metrics include R² (variance explained), RMSE (root mean squared error), MAE (mean absolute error), and MAPE (mean absolute percentage error) \cite{Botchkarev2018PerformanceM}\cite{Chicco2021TheCO}. For classification (energy-tier prediction), we report accuracy (correct predictions), precision (positive prediction accuracy), recall (true positive coverage), F1-score (harmonic mean of precision and recall), and confusion matrices \cite{SOKOLOVA2009427}. All reported results include mean and standard deviation across folds.

% \begin{figure}[t]
%   \centering
%   \includegraphics[width=0.9\linewidth]{figures/energy_distribution_by_block_type.png}
%   \caption{Energy consumption distribution across different block types.}
%   \label{fig:energy-distribution-block-type}
% \end{figure}

%% file: sections/research_question.tex
\section{Research Questions}
\label{sec:research-questions}

% Our work addresses the overarching problem of how energy efficiency can be made a design time concern for software developers.
To guide our investigation, we formulate the following three research questions. These questions are answered through the analyses presented in Section~\ref{sec:results}.

\textbf{RQ1: How can energy consumption of small code blocks be measured accurately and reproducibly?}  
We evaluate the proposed methodology's measurement granularity, stability and effectiveness for small code blocks.

\textbf{RQ2: What statistical relationships exist between static code features and block-level energy consumption?}  
We investigate whether code structure and metrics are statistically related to energy consumption. This leads to two sub-questions:
\begin{itemize}
  \item \textbf{RQ2a}: Which static code features exhibit associations with block's energy consumption?  
  \item \textbf{RQ2b}: What is the nature of these associations between energy consumption and code features?
  %   \item \textbf{RQ2a}: What kind of code aspects have high impact on block's energy consumption?  
  % \item \textbf{RQ2b}: How do structural and contextual features derived from Abstract Syntax Trees correlate to energy?
\end{itemize}
% These are answered through correlation analyses (Pearson, Spearman, Kendall) and feature importance rankings across features.

\textbf{RQ3: How reliably can code features predict block-level energy consumption?}  
Beyond correlation, we ask whether predictive models trained on static features can predict energy consumption and generalize to unseen code. This leads to two sub-questions:
\begin{itemize}
  \item \textbf{RQ3a:} How accurately can regression models predict absolute block-level energy values?  
  \item \textbf{RQ3b:} How effectively can classification models assign blocks to energy tiers (low, medium, high)?  
\end{itemize}
% These are answered through regression and classification performance metrics, cross-validation stability, and analysis of generalization gaps between training and test sets.

% These questions are answered through the implementation of our Visual Studio Code extension, which highlights energy-intensive blocks inline with source code, enabling targeted refactoring at design time.

%% file: sections/7.Results.tex
\section{Results}
\label{sec:results}

We present the results according to the research questions defined in Section ~\ref{sec:research-questions}.

\subsection{RQ1: How can energy consumption of small code blocks be measured accurately and reproducibly?}

\begin{figure}[t]
  \centering
  \includegraphics[width=1\linewidth]{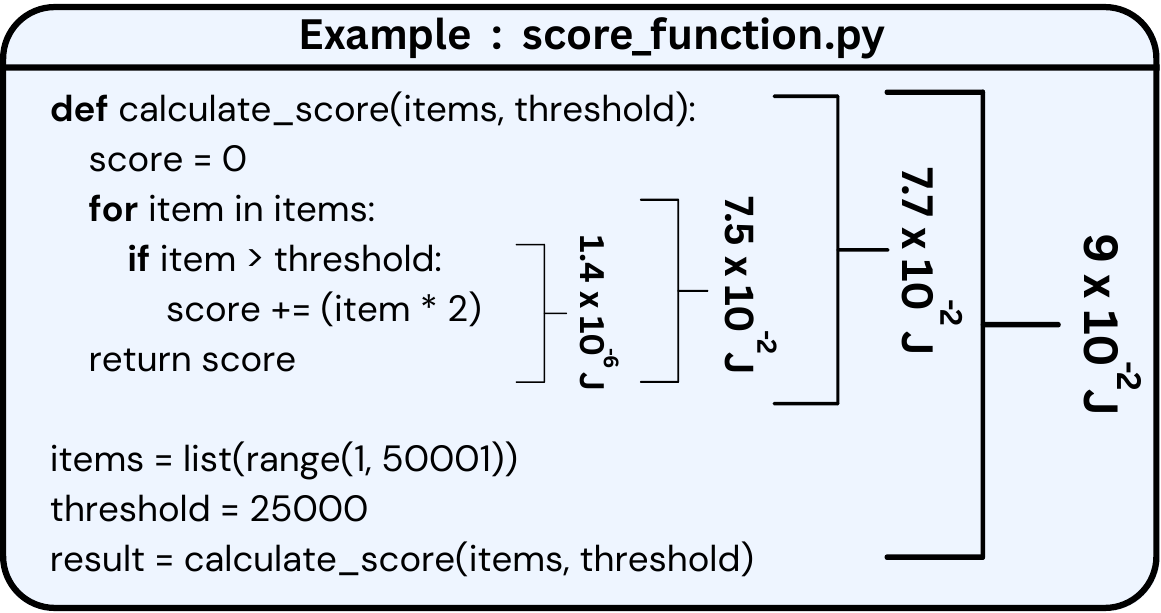}
  \caption{Block Level Energy Measurement of the score computation function}
  \label{fig:example-measurement}
\end{figure}
Our first research question examines whether energy consumption of microsecond-scale code blocks can be measured accurately and reproducibly.  Each block was measured ten times, and across the dataset \textbf{more than 90\%} of the blocks exhibited \textbf{less than 10\% variation} between repeated trials after IQR outlier removal, indicating strong measurement stability and effective suppression of environmental noise. The results demonstrate that the proposed measurement infrastructure provides stable and reliable energy values at block granularity. This is a result of system stabilization and calibrated padding loops.

The measurement protocol also achieves both high sensitivity and broad coverage. The observed energy values spans six orders of magnitude, ranging from $2.37 \times 10^{-5}$ joules for trivial blocks to $7.48 \times 10^{2}$ joules for complex functions. This wide dynamic range confirms that the methodology is capable of capturing energy signatures of extremely short-lived constructs while remaining reliable. Without the amplification and synchronization mechanisms, many of these fine-grained blocks would register highly unstable or zero readings using conventional software-profilers.

\begin{figure}[t]
  \centering
  \includegraphics[width=1\linewidth]{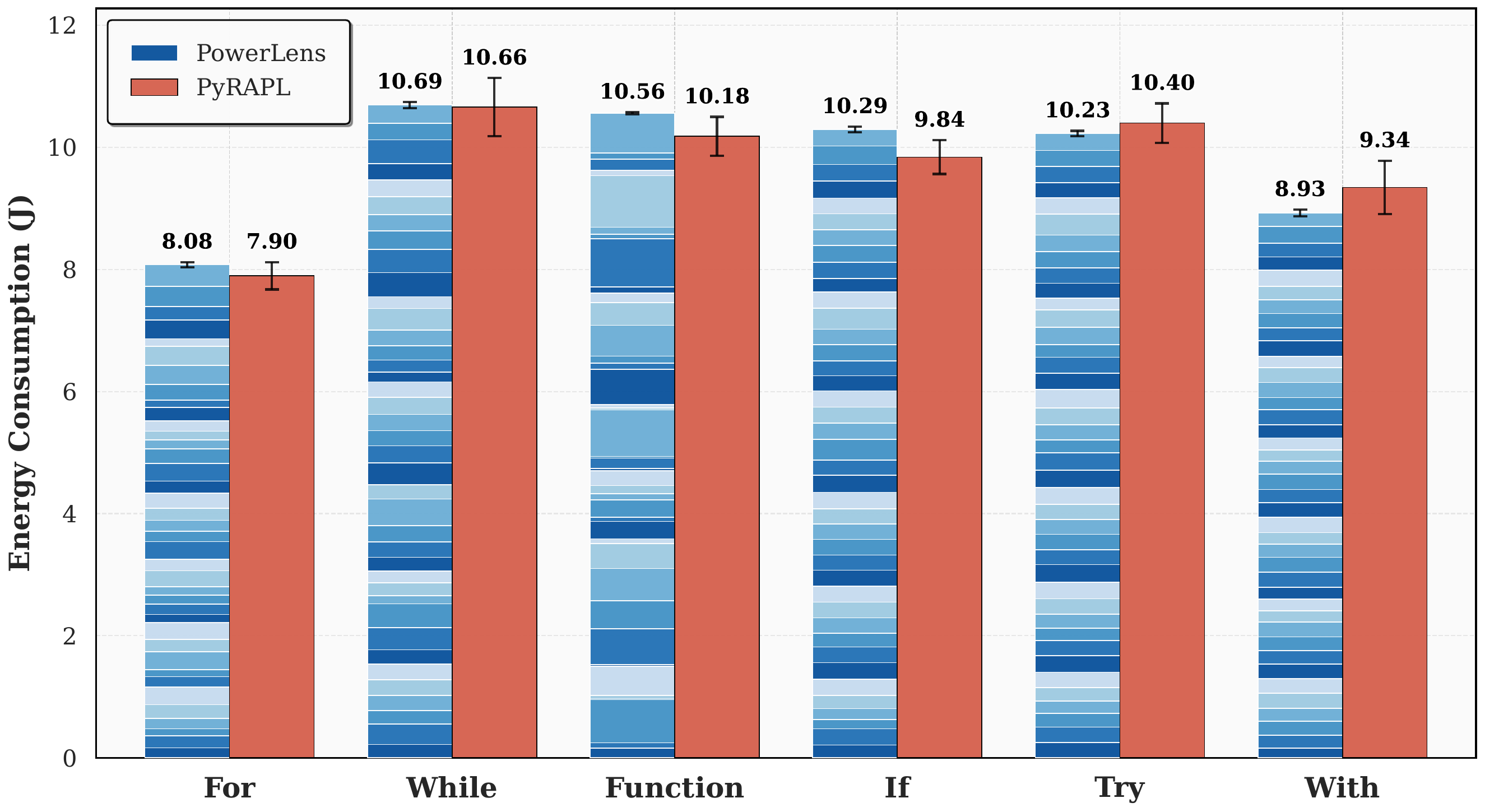}
  \caption{Sum of Blocks measured by Powerlens compared with Overall measurement by Pyrapl}
  \label{fig:powerlens-pyrapl}
\end{figure}

Figure~\ref{fig:example-measurement} illustrates block-level energy measurements for the \texttt{score\_function.py} example, where annotated values indicate per-block energy consumption in joules as measured by PowerLens in Nested Blocks.
Figure~\ref{fig:powerlens-pyrapl} compares the sum of block-level energy measurements obtained with PowerLens against coarse-grained measurements from PyRAPL\footnote{\url{https://github.com/powerapi-ng/pyRAPL}}, grouped by construct type. To \textbf{validate} the correctness of the fine-grained measurements, we compared the aggregate energy obtained by summing block-level measurements from PowerLens with the total program-level energy reported by PyRAPL for the same codes, across 40 instances of each block type (see Figure~\ref{fig:powerlens-pyrapl}). For all block categories, the aggregated PowerLens measurements closely match the corresponding coarse-grained energy reported by PyRAPL. In addition, PowerLens exhibits \textbf{substantially lower variance} across repeated measurements, as it stabilizes the execution environment and applies calibration loops to isolate block execution energy.
% These results confirm that \textbf{PowerLens} provides \textbf{reliable, high-resolution} energy measurements across diverse code constructs while remaining consistent with established coarse-grained profiling tools.

\begin{rqbox}{Answer to RQ1}Together, these results confirm that PowerLens enables accurate, stable, and reproducible energy measurement at the level of individual small code blocks, overcoming the temporal limitations of existing software-based profilers.
\end{rqbox}

% \textbf{\textit{PLACEHOLDER: Boxplot of measurement variance across blocks; histogram of block-level energy distribution.}}

\subsection{RQ2: What statistical relationships exist between static code features and block-level energy consumption?}

\input{Tables/top20_correlation_feature_importance}

We next examine whether code metrics of source code provide meaningful explanatory signals for energy consumption and the nature of these signals.

\noindent
We present the top 20 features from the correlation analysis, ranked by absolute correlation values (Pearson, Spearman, and Kendall) and feature importance scores (Extra Trees, Random Forest, and Gradient Boosting), in Table~\ref{tab:feature_ranking}.

The feature importance results in Table~\ref{tab:feature_ranking} indicate that energy consumption does not strongly depend on any single feature. While several metrics exhibit statistically meaningful correlations with energy, most Pearson correlation coefficients are moderate in magnitude. In contrast, Spearman and Kendall coefficients are substantially higher for many features, suggesting that the relation between code features and energy consumption is largely \textbf{non-linear}.

Features such as \texttt{operator\_density}, \texttt{operator\_entropy}, and \texttt{loops\_count} demonstrate linear associations and consistent predictive power across models. Notably, \texttt{operator\_density} ranks first in all three models. Other features, including \texttt{call\_density}, \texttt{conditionals\_count}, \texttt{literal\_density}, and \texttt{unique\_node\_types}, exhibit non-linear relationships with energy while maintaining stable predictive importance across models. The convergence of correlation analysis and predictive performance suggests that the models capture genuine underlying relationships.
Although some features achieve high rank-based correlations primarily because they act as categorical indicators, \texttt{functions\_count} exhibits high Spearman and Kendall coefficients but contributes little to prediction. Such features carry block specific knowledge and hence, effectively distinguish block types but can't predict energy consumption.
\begin{rqbox}{Answer to RQ2}
Taken together, these findings demonstrate that energy consumption is encoded in the compositional structure of code. Most feature-energy relationships are non-linear, as evidenced by the gap between Pearson and Spearman coefficients. Control flow shape, operator composition, functional decomposition, and nesting patterns jointly determine energy behavior. As a result, features that attempt to explain energy using isolated characteristics are inherently limited. Effective energy modeling requires holistic representations of code capturing feature 
interactions.
\end{rqbox}

\subsection{RQ3: How reliably can code features predict block-level energy consumption?}

\input{Tables/regression}
\input{Tables/classification}
We assess prediction accuracy of static features on unseen code blocks and identify effective modeling choices.

As shown in Table ~\ref{tab:regression_log}, ensemble and tree-based regression models achieve strong predictive performance, with the XGBoost regressor attaining the highest test performance ($R^2$ of 0.755). SVR and Gradient Boosting follow closely, exhibiting similarly high accuracy and stable cross-validation results. These models outperform linear models, which struggle to capture the complex feature interactions present in the data. The energy distribution in our dataset is highly skewed, spanning several orders of magnitude. To mitigate this skewness and stabilize learning, we apply square root and logarithm transformations to the target values. These transformations significantly improves model convergence and generalization.

In addition to absolute error metrics, we evaluate relative prediction quality using MAPE. While absolute errors remain small, they are less informative given that the dataset spans several orders of magnitude. MAPE is therefore more appropriate; however, it overstates relative error for extremely low-energy blocks, where even minor absolute deviations result in large relative differences. Among the evaluated models, Random Forest and KNN achieve the lowest MAPE values, whereas Gradient Boosting and XGBoost exhibit higher MAPE values exceeding 1.7. This behavior reflects the intrinsic difficulty of accurately modeling micro-scale energy values and highlights the importance of reporting both absolute and relative error when evaluating fine-grained energy predictors.

The superior performance of ensemble and tree-based models is directly explained by the findings of RQ2. Since energy behavior arises from non-linear interactions among multiple code features, models capable of learning hierarchical decision boundaries and feature compositions are fundamentally better suited for this task. Tree ensembles naturally capture such interactions through recursive partitioning, whereas linear models cannot express these relationships without extensive manual feature engineering.
While ensemble models deliver the highest accuracy, they also need higher modeling energy for inference. This requires a practical energy–accuracy tradeoff: the most accurate predictors might consume more energy themselves. However, this overhead remains negligible compared to the energy savings enabled by design-time optimization. Some models might offer minimal gains in accuracy with huge difference in energy consumption (for e.g. Gradient Boosting over XGBoost) and since the estimation is only supposed to be a guiding direction to identify hotspots, modeling energy to accuracy tradeoff is considerable. 

Beyond regression, we evaluate the ability of models to categorize code blocks into energy tiers (low, medium, high) in Table ~\ref{tab:classification}. The XGBoost classifier achieves the best performance, with 80.6\% accuracy and very stable cross-validation results. These classification results demonstrate that static features support reliable qualitative assessment of energy behavior, which is particularly valuable for developer-facing tools where actionable guidance matters.

\begin{rqbox}{Answer to RQ3}
Overall, the results demonstrate that block-level energy can be predicted accurately and robustly from static code features.  Ensemble models are effective as they capture the non-linear feature interactions inherent in energy behavior, and small
generalization gaps across folds confirm model stability. Lightweight Ensembles would be ideal in balancing modeling energy and performance.
\end{rqbox}

\subsection{Feature Group Ablation}
\label{sec:ablation}
\input{Tables/table_leave_one_group}
% \input{Tables/table_single_group}
% \begin{bluepar}
To assess the contribution of each feature category, we conducted a
systematic ablation (Table~\ref{tab:ablation}) using XGBoost under two complementary settings:
(a)~\textit{leave-one-group-out}, removing each of the groups
individually, and (b)~\textit{single-group-only}, training on each
group in isolation.
No single removal causes a dramatic performance collapse, confirming
that feature groups contribute complementary signals (Table~\ref{tab:ablation}). The largest drops occur without \textit{Density} ($-$2.8~pp) and \textit{Counts} ($-$2.6~pp).
\textit{Counts} achieves the highest standalone performance ($R^2 =
0.720$, Acc~$= 74.2\%$), yet no individual group approaches full-model
performance (Table~\ref{tab:ablation}). Most notably, \textit{Complexity} metrics alone yield $R^2 = 0.067$ and $51.2\%$ accuracy barely above random directly
confirming that conventional complexity proxies are insufficient energy
predictors.

%% file: Tables/top20_correlation_feature_importance.tex
\begin{table*}[htbp]
\centering
\caption{Top 20 features by Correlation and Feature Importance}
\resizebox{\textwidth}{!}{%
\begin{tabular}{lllllllllllll}
\hline
Rank & \multicolumn{2}{l}{Pearson ($|r|$)}          & \multicolumn{2}{l}{Spearman ($|\rho|$)}      & \multicolumn{2}{l}{Kendall (|$\tau$|)}       & \multicolumn{2}{l}{ExtraTrees}               & \multicolumn{2}{l}{Random Forest}            & \multicolumn{2}{l}{Gradient Boosting}        \\ \cline{2-13} 
     & Feature                     & Value          & Feature                     & Value          & Feature                     & Value          & Feature                     & Value          & Feature                     & Value          & Feature                     & Value          \\ \hline
1    & \textbf{operator density}   & \textbf{0.286} & \textit{functions count}    & \textit{0.621} & \textit{functions count}    & \textit{0.507} & \textbf{operator density}   & \textbf{0.086} & \textbf{operator density}   & \textbf{0.099} & \textbf{operator density}   & \textbf{0.102} \\
2    & \textbf{operator entropy}   & \textbf{0.205} & node type entropy           & 0.294          & node type entropy           & 0.193          & \textbf{loops count}        & \textbf{0.063} & \textit{unique node types}  & \textit{0.075} & \textbf{program difficulty} & \textbf{0.056} \\
3    & conditionals count          & 0.181          & \textit{conditionals count} & \textit{0.229} & \textit{conditionals count} & \textit{0.178} & \textit{functions count}    & \textit{0.061} & \textbf{program difficulty} & \textbf{0.073} & program effort              & 0.054          \\
4    & unique operators            & 0.178          & cognitive complexity        & 0.225          & cognitive complexity        & 0.165          & \textbf{operator entropy}   & \textbf{0.054} & \textit{call density}       & \textit{0.072} & \textit{variable entropy}   & \textit{0.037} \\
5    & \textit{literal density}    & \textit{0.174} & nesting complexity          & 0.215          & unique functions            & 0.165          & \textit{variable entropy}   & \textit{0.054} & \textit{variable entropy}   & \textit{0.061} & \textbf{loops count}        & \textbf{0.037} \\
6    & functions count             & 0.166          & control flow complexity     & 0.210          & nesting complexity          & 0.160          & \textit{call density}       & \textit{0.053} & variable density            & 0.061          & \textit{unique node types}  & \textit{0.031} \\
7    & \textbf{loops count}        & \textbf{0.156} & \textit{literal density}    & \textit{0.207} & control flow complexity     & 0.156          & \textbf{program difficulty} & \textbf{0.052} & leaves to nodes ratio       & 0.053          & \textit{call density}       & \textit{0.022} \\
8    & \textit{variable entropy}   & \textit{0.145} & unique functions            & 0.205          & cyclomatic complexity       & 0.150          & depth variance              & 0.046          & depth variance              & 0.052          & \textit{literal density}    & \textit{0.016} \\
9    & variable density            & 0.144          & cyclomatic complexity       & 0.203          & \textit{literal density}    & \textit{0.145} & unique variables            & 0.046          & \textit{literal density}    & \textit{0.051} & \textit{conditionals count} & \textit{0.014} \\
10   & \textbf{program difficulty} & \textbf{0.129} & vocabulary size             & 0.199          & vocabulary size             & 0.141          & \textit{unique node types}  & \textit{0.043} & program effort              & 0.042          & \textbf{operator entropy}   & \textbf{0.011} \\
11   & unique variables            & 0.122          & \textbf{operator density}   & \textbf{0.195} & \textbf{operator density}   & \textbf{0.138} & literal density             & 0.042          & \textbf{operator entropy}   & \textbf{0.037} & leaves to nodes ratio       & 0.010          \\
12   & leaves to nodes ratio       & 0.117          & \textit{unique node types}  & \textit{0.191} & \textit{call density}       & \textit{0.133} & \textit{conditionals count} & \textit{0.036} & unique variables            & 0.032          & \textit{functions count}    & \textit{0.010} \\
13   & depth variance              & 0.116          & \textit{call density}       & \textit{0.183} & \textit{unique node types}  & \textit{0.127} & variable density            & 0.035          & unique functions            & 0.030          & program length              & 0.009          \\
14   & attribute density           & 0.080          & program volume              & 0.148          & unique variables            & 0.110          & unique operators            & 0.032          & \textbf{loops count}        & \textbf{0.030} & depth variance              & 0.009          \\
15   & max branching factor        & 0.073          & \textit{variable entropy}   & \textit{0.139} & \textit{variable entropy}   & \textit{0.105} & unique functions            & 0.027          & total nodes                 & 0.028          & unique operators            & 0.008          \\
% 16   & max depth                   & 0.072          & unique variables            & 0.138          & unique operators            & 0.103          & avg depth                   & 0.026          & avg depth                   & 0.026          & variable density            & 0.007          \\
% 17   & avg depth                   & 0.071          & unique operators            & 0.138          & program volume              & 0.101          & avg branching factor        & 0.025          & node type entropy           & 0.025          & unique functions            & 0.005          \\
% 18   & cognitive complexity        & 0.055          & program length              & 0.126          & \textbf{operator entropy}   & \textbf{0.094} & leaves to nodes ratio       & 0.023          & unique operators            & 0.024          & program volume              & 0.004          \\
% 19   & node type entropy           & 0.048          & \textbf{operator entropy}   & \textbf{0.125} & program length              & 0.087          & max depth                   & 0.022          & program volume              & 0.023          & unique variables            & 0.004          \\
% 20   & \textit{unique node types}  & \textit{0.047} & leaves to nodes ratio       & 0.118          & leaves to nodes ratio       & 0.080          & cognitive complexity        & 0.021          & program length              & 0.022          & node type entropy           & 0.003          \\
\hline
\end{tabular}
}\footnotesize{ \textbf{Bold} = linear relation, \textit{italic} = non-linear relation}
\label{tab:feature_ranking}
\end{table*}

%% file: Tables/regression.tex
\begin{table}[h]
\centering
\caption{Regression Models with \textbf{log} Transform on Target}
\resizebox{\columnwidth}{!}{%
\begin{tabular}{llcllll}
\hline
Model             & \multicolumn{1}{c}{Test $R^2$}            & CV $R^2$ ($\pm$ std)                                  & \multicolumn{1}{c}{RMSE}                  & \multicolumn{1}{c}{MAE}                   & \multicolumn{1}{c}{MAPE}                   & Energy (mJ)                               \\ \hline
XGBoost           & \cellcolor{gray!45}0.755 & \cellcolor{gray!45}0.811 $\pm$ 0.040 & \cellcolor{gray!45}0.281 & \cellcolor{gray!45}0.057 & 172.45                                     & \cellcolor{gray!15}15.26 \\
SVR               & \cellcolor{gray!30}0.752 & \cellcolor{gray!15}0.803 $\pm$ 0.039 & \cellcolor{gray!30}0.283 & 0.091                                    & 182.27                                     & 15.47                                     \\
Gradient Boosting & \cellcolor{gray!15}0.747 & \cellcolor{gray!30}0.810 $\pm$ 0.040 & \cellcolor{gray!15}0.286 & \cellcolor{gray!15}0.058 & 171.83                                     & \cellcolor{gray!45}13.56 \\
CatBoost          & 0.722                                     & 0.799 $\pm$ 0.043                                     & 0.300                                     & \cellcolor{gray!30}0.058 & \cellcolor{gray!15}166.35 & 22.97                                     \\
Random Forest     & 0.719                                     & 0.793 $\pm$ 0.047                                     & 0.302                                     & 0.060                                     & \cellcolor{gray!30}162.83 & 258.01                                    \\
Extra Trees       & 0.682                                     & 0.737 $\pm$ 0.047                                     & 0.321                                     & 0.074                                     & 170.85                                     & 275.83                                    \\
Decision Tree     & 0.648                                     & 0.722 $\pm$ 0.070                                     & 0.338                                     & 0.067                                     & 170.24                                     & \cellcolor{gray!30}14.69 \\
KNN               & 0.645                                     & 0.684 $\pm$ 0.081                                     & 0.340                                     & 0.066                                     & \cellcolor{gray!45}105.78 & 19.59                                     \\
AdaBoost          & 0.633                                     & 0.757 $\pm$ 0.051                                     & 0.345                                     & 0.092                                     & 171.81                                     & 20.04                                     \\ \hline
\end{tabular}
}\footnotesize{ Shading indicates top two models by each metric.}
\label{tab:regression_log}
\end{table}

% \begin{table}[h]
% \centering
% \caption{Regression Models with \textbf{log} Transformation on Target}
% \resizebox{\columnwidth}{!}{%
% \begin{tabular}{l c c c c c}
% \hline
% Model & Test $R^2$ & CV $R^2$ ($\pm$ std) & RMSE & MAE & Energy (mJ) \\
% \hline
% Gradient Boosting & \cellcolor{gray!40}0.877 & \cellcolor{gray!25}0.789 $\pm$ 0.088& \cellcolor{gray!40}0.160 & 0.038 & \cellcolor{gray!40}13.56 \\
% SVR & \cellcolor{gray!25}0.859 & \cellcolor{gray!12}0.768 $\pm$ 0.095& \cellcolor{gray!25}0.171 & 0.074 & 15.47 \\
% Random Forest & \cellcolor{gray!12}0.851 & 0.773 $\pm$ 0.105 & \cellcolor{gray!12}0.176& 0.037 & 258.01 \\
% Extra Trees & 0.845 & 0.775 $\pm$ 0.098 & 0.180 & 0.039 & 275.83 \\
% CatBoost & 0.837 & \cellcolor{gray!40}0.795 $\pm$ 0.088 & 0.184 & \cellcolor{gray!40}0.033 & 22.97 \\
% XGBoost & 0.829 & 0.770 $\pm$ 0.071 & 0.189 & \cellcolor{gray!25}0.035 & \cellcolor{gray!12}15.26\\
% Decision Tree & 0.808 & 0.708 $\pm$ 0.082 & 0.200 & 0.037 & \cellcolor{gray!25}14.69 \\
% AdaBoost & 0.747 & 0.650 $\pm$ 0.098 & 0.230 & 0.075 & 20.04 \\
% KNN & 0.748 & 0.736 $\pm$ 0.084 & 0.230 & \cellcolor{gray!25}0.035 & 19.59 \\
% \hline
% \end{tabular}%
% }
% \label{tab:regression_log}
% \end{table}

%% file: Tables/classification.tex
\begin{table}[h]
\centering
\caption{Energy Tier Classification Models}
\resizebox{\columnwidth}{!}{%
\begin{tabular}{llcllll}
\hline
Model               & \multicolumn{1}{c}{Accuracy} & CV Accuracy       & \multicolumn{1}{c}{Precision} & \multicolumn{1}{c}{Recall} & \multicolumn{1}{c}{F1} & Energy(J) \\ \hline
XGBoost             & \cellcolor{gray!45}0.806                        & \cellcolor{gray!30}0.793 $\pm$ 0.007 & \cellcolor{gray!45}0.804                         & \cellcolor{gray!45}0.806                      &\cellcolor{gray!45} 0.805                  & 0.022     \\
Random Forest       &\cellcolor{gray!30} 0.792                        & \cellcolor{gray!15}0.780 $\pm$ 0.007 & \cellcolor{gray!30}0.789                         & \cellcolor{gray!30}0.792                      & 0.789                  & 0.373     \\
Gradient Boosting   & \cellcolor{gray!15}0.788                        & \cellcolor{gray!45}0.794 $\pm$ 0.008 & \cellcolor{gray!15}0.788                         & \cellcolor{gray!15}0.788                      & \cellcolor{gray!30}0.788                  &\cellcolor{gray!30} 0.015     \\
K-NN                & 0.783                        & 0.771 $\pm$ 0.005 & 0.780                         & 0.783                      & 0.780                  & 0.027     \\
SVM                 & 0.781                        & 0.771 $\pm$ 0.009 & 0.780                         & 0.781                      & 0.778                  & 0.022     \\
Extra Trees         & 0.769                        & 0.765 $\pm$ 0.003 & 0.768                         & 0.769                      & 0.765                  & 0.315     \\
Decision Tree       & 0.749                        & 0.736 $\pm$ 0.009 & 0.744                         & 0.749                      & 0.745                  & \cellcolor{gray!45}0.014     \\
Logistic Regression & 0.735                        & 0.726 $\pm$ 0.003 & 0.729                         & 0.735                      & 0.731                  & \cellcolor{gray!15}0.017     \\
SGD Classifier      & 0.729                        & 0.713 $\pm$ 0.005 & 0.722                         & 0.729                      & 0.721                  & 0.022     \\
% QDA                 & 0.720                        & 0.709 $\pm$ 0.010 & 0.745                         & 0.720                      & 0.711                  & 0.052     \\
% LDA                 & 0.717                        & 0.712 $\pm$ 0.005 & 0.710                         & 0.717                      & 0.708                  & 0.017     \\
% Ridge Classifier    & 0.712                        & 0.706 $\pm$ 0.006 & 0.708                         & 0.712                      & 0.700                  & 0.016     \\
% Gaussian NB         & 0.678                        & 0.653 $\pm$ 0.002 & 0.680                         & 0.678                      & 0.669                  & 0.022     \\
\hline
\end{tabular}
}\footnotesize{ Shading indicates top two models by each metric.}
\label{tab:classification}
\end{table}

%% file: Tables/table_leave_one_group.tex
% \begin{table}[t]
% \centering
% \caption{Leave-one-group-out ablation. Each row removes one feature group. $\Delta$ values are
%          relative to the full model.}
% \label{tab:ablation_lopo}
% \small
% \begin{tabular}{lrrrrrl}
% \hline
% \textbf{Removed Group}     & \textbf{\#Feat} & \textbf{Test $R^2$} & \textbf{$\Delta R^2$} & \textbf{Test Acc} & \textbf{$\Delta$Acc} & \textbf{} \\ \hline
% \textit{None (Full model)} & 33              & 0.757               & ---                   & 80.6\%            & ---                  &           \\ \hline
% Density                    & 28              & 0.755               & $-$0.002              & 77.8\%            & $-$2.8 pp       &           \\
% Counts                     & 25              & 0.746               & $-$0.011              & 78.0\%            & $-$2.6 pp       &           \\
% Halstead                   & 28              & 0.752               & $-$0.006              & 80.4\%            & $-$0.1 pp       &           \\
% Complexity                 & 29              & 0.755               & $-$0.002              & 80.1\%            & $-$0.4 pp       &           \\
% Entropy                    & 30              & 0.758               & $+$0.001              & 80.5\%            & $-$0.1 pp       &           \\
% AST Structural             & 25              & 0.757               & $\sim$0               & 80.4\%            & $-$0.1 pp       &           \\ \hline
% \end{tabular}
% \end{table}

% \begin{bluepar}
    
\begin{table}[t]
\centering
\caption{Feature group Ablations}
\label{tab:ablation}
\small
\begin{tabular}{lrr rrr}
\toprule
& \multicolumn{2}{c}{\textit{Leave-One-Out}} 
& \multicolumn{3}{c}{\textit{Group Only}} \\
\cmidrule(lr){2-3} \cmidrule(lr){4-6}
\textbf{Group} & $\Delta R^2$ & $\Delta$\textbf{Acc} 
               & \textbf{\#Feat} & $R^2$ & \textbf{Acc} \\
\midrule
Density        & $-$0.002 & $-$2.8~pp & 5 & 0.700 & 71.1\% \\
Counts         & $-$0.011 & $-$2.6~pp & 8 & 0.720 & 74.2\% \\
Halstead       & $-$0.006 & $-$0.1~pp & 5 & 0.688 & 59.7\% \\
Complexity     & $-$0.002 & $-$0.4~pp & 4 & 0.067 & 51.2\% \\
Entropy        & $+$0.001 & $-$0.1~pp & 3 & 0.691 & 68.3\% \\
AST Structural & $\sim$0  & $-$0.1~pp & 8 & 0.650 & 69.4\% \\
\bottomrule
\end{tabular}
\end{table}
% \end{bluepar}

%% file: sections/8.Discussion.tex
\section{Discussion}
\label{sec:Discussion}
\subsection{Lessons Learned}

\textbf{Energy can now be treated as a metric of code Quality.} Several long-standing challenges identified in prior survey and vision work on energy-aware software engineering can be addressed through fine-grained, design-time energy guidance \cite{LEE2024111944}. The results for \textbf{RQ1} and \textbf{RQ3} demonstrate that energy consumption can be measured and estimated meaningfully at the level of small code blocks; this granularity aligns with how developers reason about and refactor code. In doing so, our methodology directly responds to earlier calls for improving energy observability beyond application and method-level measurements\cite{10.1145/2425248.2425252,7429297}. PowerLens reduces reliance on external hardware power meters to obtain high-resolution measurements. 

\noindent
\textbf{No single metric is a proxy for Energy.} Contrary to approaches that treat some software-quality metrics (for example, size or cyclomatic complexity) as proxies for energy\cite{10329262}, our findings for \textbf{RQ2} indicate that no single metric explains energy consumption to a large extent. Rather, energy behavior emerges from interactions among multiple code properties. The observed correlations and feature importance between block-level energy consumption and code metrics reinforce the view that energy is encoded in the compositional code features rather than isolated metrics.
The results for \textbf{RQ2} and \textbf{RQ3} suggest that energy efficient ensemble models should be used to provide actionable energy guidance. These simple machine learning approaches provide approximate, design-time estimates which can be sufficient to distinguish energy-efficient constructs from energy-intensive ones. This aligns with survey studies that emphasize the practical value of timely but approximate estimation over delayed precise measurements\cite{ZARAGOZA2025115889,LEE2024111944}. 

\subsection{Implications for Practice}
% \begin{bluepar}
From a practitioner's perspective, the key implication of this work is that energy efficiency can be considered meaningfully during code construction rather than relegated to late-stage profiling. EnCoDe's primary goal is relative hotspot ranking within a codebase, not absolute Joule-level accuracy. The classifier flags blocks predicted as \textit{High} energy tier with a lint-like warning. The developer examines the flagged block and considers alternative implementations. The regression model provides a comparative estimate between the original and refactored version to guide the choice. This workflow mirrors how developers already use static analyzers \cite{su14148596}.
% \end{bluepar}
Energy-tier classification  evaluated in \textbf{RQ3} further lowers the barrier to adoption. The system surfaces lint-like information that points to constructs likely to be energy-intensive. This mirrors successful adoption patterns for other static analyzers and suggests energy concerns can be integrated into routine development practices. Embedding energy awareness in the development lifecycle also aligns with emerging SusDevOps perspectives, where sustainability considerations are addressed alongside performance, reliability, and maintainability \cite{11023955,10.1145/3680470}. Our empirical results provide initial evidence that such integration is feasible in practice.

\subsection{Implications for Research}

For researchers, this work demonstrates how open challenges in energy-aware software engineering can be addressed via an integrated empirical and static approach \cite{LEE2024111944,ZARAGOZA2025115889,electronics14071331}. By coupling fine-grained runtime measurements with static code features, our methodology operationalizes calls to combine runtime energy data and design-time artifacts across the software lifecycle. This combined methodology enables several promising research directions: large-scale mining studies across languages and ecosystems, comparative analyzes of energy behavior across programming paradigms and systematic investigations into energy-aware refactoring.

Beyond methodological advances, our findings lend empirical support to treating energy consumption as a first-class software quality attribute. Prior work has argued conceptually for elevating energy alongside traditional qualities such as performance and maintainability; our fine-grained evidence helps close the gap by showing how energy can be measured, modeled, and reasoned about at the level of individual constructs. Finally, the availability of empirically grounded, block-level datasets opens opportunities to study energy behavior in emerging domains, such as AI-intensive systems and other compute-heavy applications, where understanding fine-grained energy signatures is increasingly critical.

%% file: sections/9.Conclusion.tex
\section{Threats to Validity}
\label{sec:threats}
Our study is bounded by some validity concerns which we discuss following the established guidelines~\cite{10.5555/2349018}.

\textbf{Internal validity.}  
Data leakage was minimized by strict train–test separation and stratified 5-fold cross validation was employed. Amplification and Calibration steps assume linear scaling of energy consumption. Overfitting was monitored via cross-validation, with small observed gaps suggesting good generalization. For stability, execution was pinned to a single CPU core, but this limits our ability to capture multi-threading effects.

\textbf{Construct validity.}  
Our methodology does not model how energy scales with different user inputs, data sizes, or runtime states. Amplification factor (N=1000) may not be ideal for ultra small blocks.  RAPL readings are limited to CPU and DRAM , excluding I/O, network, or storage. Further, RAPL readings have inherent measurement errors. Our methodology does not model inter-thread interactions or distributed execution patterns.

\textbf{External validity.}  
 Though the dataset is representative, it does not cover the full spectrum of libraries, code styles, or domain specific. All experiments were conducted on a single CPU but energy characteristics may differ on other processors. Validation across languages and hardware is necessary to establish generality.

% \textbf{Granularity and scope.}  
% Our measurement protocol pins execution to a single CPU core to reduce scheduling noise and ensure consistent readings. While this improves stability, it also forces workloads into single-core execution, limiting our ability to capture the effects of multi-threading or multiprocessing. More broadly, our analysis is confined to block-level constructs such as functions, loops, and conditionals. We do not capture higher-level program behaviors such as inter-thread interactions, distributed execution, or application-level optimization strategies.

% \medskip
% In summary, our work establishes the feasibility of block-level energy prediction under controlled conditions. However, limitations in dataset coverage, hardware diversity, runtime modeling, and system-level energy attribution highlight the need for further research before generalizing to all software engineering contexts.

\section{Conclusion and Future Work}
\label{sec:conclusion}
This paper introduced \textsc{EnCoDe}, a methodology for design-time, fine-grained energy estimation of code blocks, together with \textsc{PowerLens}. PowerLens enables consistent measurement of sub millisecond-scale code blocks. By shifting energy awareness from late-stage profiling to early design, this work supports treating energy as a first-class software quality attribute. A natural next step is to extend beyond Python to other programming languages, runtime environments and modelling techniques, enabling validation across programming paradigms and software stacks. Beyond traditional applications, the same block and component-level analysis can be applied to LLMs and machine learning systems, which consume a lot of energy. Moving from detection to actionable guidance and automated repair will make sustainability effective and accessible.  

% Looking further ahead, Packaging \textsc{EnCoDe} as IDE-integrated tools and CI checks can make energy feedback as routine as performance or code-quality warnings, while block-level feedback can support hands-on teaching of green software engineering. At a larger scale, PowerLens datasets and structural insights can inform programming assistants and large language models to generate energy-efficient code and reason about energy trade-offs alongside correctness and performance, helping establish energy efficiency as a first-class software concern.\section{Conclusion and Future Work}